# Can quantum computing solve classically unsolvable problems?

Andrew Hodges

Wadham College
University of Oxford
Oxford OX1 3PN, U.K.
andrew.hodges@wadh.ox.ac.uk

**Abstract:** T. D. Kieu has claimed that a quantum computing procedure can solve a classically unsolvable problem. Recent work of W. D. Smith has shown that Kieu's central mathematical claim cannot be sustained. Here, a more general critique is given of Kieu's proposal and some suggestions are made regarding the Church-Turing thesis.

**0. Introduction:** In 2001 the physicist Tien D. Kieu announced that a quantum computing procedure could solve a classically unsolvable problem, namely Hilbert's Tenth Problem. This problem, that of deciding whether a polynomial integer-valued function of integers ever vanishes, is essentially equivalent to the standard 'halting problem' and other such unsolvable problems as defined by Turing machines. (The classic exposition of this fascinating connection is by Martin Davis (1958, 1982)). Kieu has since produced a number of publications with overlapping material. Reference will be made in what follows to some of the arXiv versions of his exposition. But it should be noted that his work has also been accepted and published in leading journals.

Kieu's claim is closely connected with the recent programme of so-called 'hypercomputation' as laid out by Copeland, Stannett, Ord, Calude and others (for a recent review see (Stannett 2004)). According to this view, as expounded in *Scientific American* by the philosophers Copeland and Proudfoot (1999), the computing of classically uncomputable functions ('hypercomputing') may be a matter of finding the right technology to make it possible. Indeed these authors enthusiastically announced the search for such technology to be already under way. Kieu's proposal emerged as perhaps the brightest hope for fulfilling this dream. It is also the strongest claim made for power of quantum computation. These questions are frequently expressed in terms of the *Church-Turing thesis,* but for reasons that will become clear, we are postponing a discussion of this topic to §8 below.

Kieu's proposal is essentially a brute-force search, in which the polynomial is evaluated for all possible arguments and the results scanned for a zero. Clearly, a search using classical computation would never decide in a finite time that a polynomial was non-vanishing. The idea is that quantum computation enables an essential speed-up, making a decision in finite time possible, because all the computations can be done in parallel, using the superposition of states in quantum mechanics. However, Kieu does not use standard quantum computation, based on 'qubits' and elementary 'gate' operations, but the 'adiabatic cooling' method due to Farhi et al. (2000). For finite searches this method yields the same quadratic speed-up as the standard qubit-based method due to Grover (1996). Kieu's claim is that for the infinite search required in the Hilbert problem, the speed-up can actually turn an infinite time into a finite time. This is of course a surprising claim, as it is stronger than an exponential speed-up. The central idea is quite ingenious: the polynomial itself is used to define a Hamiltonian for the dynamical quantum process, and the claim is that this yields the required speed-up. The adiabatic method is claimed to work by detecting (through a complicated procedure involving quantum measurement) the *least* value achieved by the square of the polynomial. One then sees whether or not it is zero.

There are some gaps in explaining exactly what the procedure is, and serious problems with the mathematical analysis of the physical system in Kieu's work. Conjectures rather than proofs are given, especially in connection with the properties of infinite-dimensional Hilbert spaces. The central point is that Kieu's procedure depends on the truth of a result which is very much stronger than the adiabatic theorem. This result is not proved by Kieu, and Warren D. Smith has found counter-examples which disprove it. This paper will not discuss these more technical problems, for which the reader is referred to Smith (2005). It surveys some of the general questions raised by Kieu's scheme that would still be problematic even if the mathematical argument could be in some way rectified. It is based on an earlier, preliminary note (Hodges 2004).

There are quite a number of such general questions to discuss, because Kieu's papers are not limited to describing this quantum procedure. They also offer further reasons for why, in his opinion, this procedure outdoes classical computation.

1. **Randomness and computability.** Important in the theoretical background is Kieu's assertion (Kieu 2003d, p. 18) that the *randomness* in his scheme takes it out of the arena of classical computability and its limitations.

A random real number is (with probability 1) uncomputable — as follows from simple cardinality arguments. Some proponents of 'hyper-computation' consider this is a key fact leading towards their goal. Yet intuitively, it seems implausible that by making some elements in a process random rather than deterministic, we would *increase* the power of that process. As Ord (2002) puts it, can this uncomputability be *harnessed* in any way? This question was addressed by Shannon et al. (1956). Their conclusions can be summarised thus: introducing a random element producing equiprobable outputs of 0 and 1 will, indeed, do no better than a deterministic Turing machine. If the probability is not 1/2 but some other computable parameter, the situation is the same. If the probability is $\lambda$, some *uncomputable real number,* then that number $\lambda$ can be computed; in other words, computation is indeed strictly extended to become computation relative to $\lambda$. Kieu (2003d) suggests that this extension is the fundamental reason why his procedure is effective. However, it must be noted that the randomness in the output can do no better than *copying* the uncomputable number that has already been put into the random element. It does not *create* anything uncomputable. We shall return to this point at the end of §4.

It is not just that Kieu has randomness in the procedure; there is also randomness in the decision effected by the procedure. The answer is only asserted to be correct with some probability. Kieu calls this a 'stochastic' version of the decision problem. It could be objected that such a 'stochastic' solution is not a solution at all: Hilbert calls for a decision, not a guess! Kieu (2004) quotes Turing on how the classical arguments of undecidability do not apply if 'mistakes' are allowed, and the word 'mistakes' should sound a warning, since the essence of the decision problem as originally posed is that we should be guaranteed a correct answer. One natural demand on any 'stochastic' version of the decision procedure, for it to count as a candidate for serious consideration, is that it should satisfy a strict uniformity constraint. For any ε >0, there should be a specified procedure P(ε) which can be applied to any polynomial, and will give the right answer with probability 1 – ε. Kieu's scheme does appear to cater for such uniformity, although it is not explicitly stated as an essential criterion. However, it may be objected that this still does not match the *certainty* of a classical decision procedure.

2. **Randomness in quantum mechanics.** There are two essentially different fundamental processes in quantum mechanics. One is the *unitary evolution* of a wave-function in configuration space. This is deterministic, governed by the Schrödinger equation with a particular Hamiltonian, and produces complex-valued amplitudes. The second is the *reduction or measurement* process which 'collapses' the wave function to an eigenstate of the measurement operator. This is where randomness enters: the probability of collapsing to any particular eigenstate is given by the squared modulus of the corresponding component of the amplitude.

The 'Heisenberg Uncertainty Principle' is often referred to as basic to quantum mechanics, but it is better considered as derived from these more fundamental processes. This has the effect of locating the randomness more clearly and such clarity is important because, in Kieu's theory, randomness is supposed to be the source of the power of the method.

Randomness plays a role in Kieu's theory because the procedure requires detecting when a particular property is satisfied by the quantum amplitude. But the amplitudes cannot be inspected directly; they can only be probed by measurements which collapse the wave-function and so introduce randomness. Kieu's idea turns on estimating the property of the amplitude by doing such measurements repeatedly, then using standard statistical theory to estimate the property of the amplitude to any required degree of certainty. Smith (2005) comments that this statistical analysis is not actually given by Kieu, but that it can be supplied. However, to achieve *complete* certainty we would need an infinite sample, which would take an infinite time, thus rendering the whole exercise futile. As mentioned above, it could be argued that this point alone renders the scheme invalid: there seems no reason why this infinite time should be considered any less problematic than the infinite time needed in classical computing.

3. **The computability of standard quantum computing.** The basic idea of standard quantum computing is that a sequence of unitary operations is applied to an initial quantum state which contains the input data. Only at the end is a reduction or measurement process applied and the output read. The unitary evolution is itself deterministic and therefore could in principle be simulated by a classical computer. The point is that this classical simulation may be essentially slower, so that quantum computing can improve on the complexity of the classical procedure. But this argument indicates that quantum computing cannot produce classically *uncomputable* results.

How then does Kieu explain the allegedly superior power of his method? Kieu (2003a, p. 9, and again in 2003c p. 27) states that his procedure does better than standard quantum computing with qubits. And he gives the reason for this superiority: it lies in the infinite dimensionality of the Hilbert spaces he uses, as opposed to the finite dimensionality of standard quantum computing. He does not emphasise that this means that he is charging a simple and finite physical system (a harmonic oscillator) with an *infinite amount of data.* But this is a fundamental point: the basis of Kieu's claim is that an infinite amount of information can be compressed into a finite physical system.

Possibly, the contrast with standard quantum computing is obscured by the way a 'qubit' is described in standard accounts. A 'qubit' is generally said to be a particular and exact superposition of $|0\rangle$ and $|1\rangle$ states, i.e. a particular ray in the complex projective space of states. It might appear therefore that a qubit needs infinite precision in its specification. Indeed Kieu (2003b, p.3) seems to imply this when he asserts that real numbers, as opposed to integers, are needed to define quantum states. But this overlooks the fact that the properties of quantum computing are always discussed relative to *bounded error.* Error and the sophisticated control of error are central elements in the development of quantum computing theory. In contrast, Kieu's infinite data storage needs *zero error* to work, and really does depend on setting up and maintaining an infinitely precise system.

**4. Infinite precision.** The basic difficulty here is not essentially quantum-mechanical. It arises in just the same way if we consider encoding an infinite amount of data into a real-valued continuous curve on [0,1] through Fourier decomposition. One may imagine a wave on a string with unboundedly high harmonics, with the amplitude of the $N$ th harmonic encoding the $N$ th piece of data. Unboundedly high harmonics correspond to unboundedly short wavelengths, so that detecting *every* harmonic requires infinite precision. Measurement apparatus incapable of a resolution better than wavelength $1/N$, will be deaf to harmonics higher than the $N$ th, and lose all the data after the $N$ th piece.

The quantum harmonic oscillator employed in Kieu's scheme has Hermite eigenstates whose wave-functions are are not essentially different in this respect from the sinusoidal functions of elementary Fourier analysis. The $N$th wave-function consists of $N$ peaks and troughs packed into the characteristic length $L$ of the oscillator. The ability to compress unboundedly many such peaks into a finite length $L$ is essential to Kieu's procedure.

The diophantine equations we are considering are equations in *many* integer variables, so the search is actually a search for a vector of several integers rather than for one integer, but this makes no essential difference to the point being made here, and for brevity we will refer in what follows to single integer arguments. We now consider what is involved in simulating the Kieu process on a classical computer. Normally, one would appeal to the idea of getting a better and better approximation in the simulation of the Schrödinger equation by taking shorter and shorter step-lengths. But this is not a 'normal' process. There is no prior knowledge of where the least value of the polynomial will occur: the process must involve *all* integer values. But taking a step-length of $1/N$ is, roughly, equivalent to computing the polynomial only up to $N$. This is *no approximation at all* to solving Hilbert's Tenth Problem, and anyway is obviously no better than what could be done much more simply by computing the polynomial. Thus, infinite precision is crucial.

Here are two more illustrations of why infinite precision is essential to Kieu's procedure.

(1) The initial step in the procedure consists of mapping the polynomial

into a physical Hamiltonian. But whilst the polynomial is defined on the integers, the Hamiltonian (with the dimensions of energy) must take values in the (real) continuum. (Indeed it is essential to the adiabatic method that it is multiplied by a continuously varying real time parameter.) As a concrete example, take the equation $2M^2 - N^2 = 0$. This obviously has no solution since $\sqrt{2}$ is irrational. According to Kieu, the quantum procedure will reveal this fact — not by the elegant argument known to the ancient Greeks, but by the infinite brute force evaluation of $(2M^2 - N^2)^2$ for all integers, and finding that its least value is 1. The first step is to encode this function as a Hamiltonian, but this means translating the 2 as 2.00000... and the 1 by 1.00000... The slightest error in this transcription will (for sufficiently large values of $M$ and $N$) completely wreck the calculation of the polynomial and invalidate the search for its minimum value.

(2) The procedure requires application of the 'number operator' as a measurement process. In the usual formalism of quantum mechanics this operator is written as $a^+ a$ and gives the impression that everything takes place in an integer-valued algebraic setting. However, the fact is that the number operator is none other than:

$$-L^2 \frac{\partial^2}{\partial x^2} + \frac{x^2}{L^2}$$

where $L$ is the characteristic length (a real-valued constant) of the oscillator. Applying it requires knowledge of the characteristic length $L$ to infinite precision.

Kieu is aware of the unboundedly short wavelengths which prevent simulation by classical computation. But Kieu (2003d, p.5) asserts that 'the probabilistic nature of the method' is manifested in this phenomenon. If this means anything, it is a claim that the errors involved in using numerical computation with non-zero step-lengths can be considered as *random* errors, to be identified with the randomness which is the alleged source of uncomputable output. This is unjustifiable. (1) There is no reason whatever why errors due to an *approximate* classical computation should produce miraculous effects *outdoing* classical computation. (2) As we noted, randomness appears in quantum mechanics only at the *reduction* stage, not during the unitary evolution stage which is being simulated by the computation.

It is worth looking again at the theorem of Shannon et al. at this point. It shows that uncomputable results will not result from the measurement process unless the values of the quantum amplitudes are uncomputable. (This condition again illustrates the necessity for infinite precision in the physical process, since for a real number to be uncomputable it must be specified with infinite precision). But the amplitudes arise completely deterministically from the initial conditions. Thus quantum-mechanical randomness cannot *in itself* be the source of any 'hypercomputing' power. This confirms that it is not the randomness but the infinite precision which is the source of any power there is in Kieu's scheme. In fact, Kieu's identification of randomness as the source of 'hypercomputing' power seems to be entirely spurious.

5. **Accelerated machines and unboundedly small time-steps.** With an elementary use of geometric progressions, it is easy to show that unsolvable problems can be solved in a finite time by Turing machines, provided their steps can be made of unboundedly short time-length. This condition is, of course, completely unphysical, but there is some mathematical interest in the definition of such 'accelerated Turing machines' (and indeed no less a figure than Hermann Weyl started this line of thought). Even more interesting, the work of W. D. Smith shows how such infinite miniaturisation in space and time can be derived from actual equations of physical interest, e.g. the Navier-Stokes equations for hydrodynamics (Smith 2003), provided these equations are taken as valid at all scales. (Obviously, this condition is quite unphysical, but the results still hold great interest). Kieu's use of unboundedly short wavelengths might perhaps be considered as playing a role similar to Smith's conditions. For this reason his analysis, if it can be corrected, may have useful content as a mathematical exploration of different kinds of computational process. Indeed W. D. Smith (2005) has demonstrated some results arising from such an investigation.

However, the unboundedly high frequencies required for Kieu's scheme, being equivalent to unboundedly short time-steps, mean that a final 'finite time' cannot be compared with the finite time taken by a classical Turing machine. For if there is no smallest unit of time, a 'finite' time cannot be compared with the finite number of unit steps taken by a Turing machine coming to a conclusion. Equivalently, if a Turing machine is allowed to have unboundedly fast steps then it too can also 'solve' unsolvable problems in a finite time.

6. **The nature of mathematical modelling**. The foregoing observations suggest a more general point about the mathematical modelling of physical systems, which is not limited to quantum mechanics, but would apply to other putative physical schemes claiming to produce classically uncomputable results. First note that if a real number is used to represent a physical *magnitude*, as is normally the case, the digits of its decimal expansion obviously decline in significance throughout the infinite sequence. But if it is the *computability* of a real number that is the subject of attention, the situation is the reverse: any finite number of digits may be changed without affecting its status, and it is always the infinite tail of remaining digits that holds the information. When studying a physical system, analogous considerations obtain. Normally, in the mathematical model of a system, perfection is not essential. Let us take, for instance, the hydrogen atom. It was a tremendous achievement of quantum mechanics to explain its spectral lines by the non-relativistic Schrödinger equation. This explanation remains worthwhile even though a more accurate relativistic equation should be used. This in turn is inadequate to explain the phenomena of full relativistic quantum field theory (e.g. the Lamb shift), and this again is known to be inadequate because of its neglect of curved space-time. In a complete theory (which we do not possess) it would be necessary to take into account the complete geometry of the universe to give a precise calculation. But the simpler models still have great value: we know that their physical predictions are reasonably *robust* against the modifications induced by more accurate theories. The situation is entirely different with a scheme such as Kieu's where total perfection of the model is required from the outset. Our usual assumptions about what is significant are turned upside-down. Indeed it is really not valid to use the normal language of 'models' in mathematical physics when setting out a scheme such as Kieu's. For such normal language involves assumptions about robustness under small changes and perturbations, and these assumptions are completely invalid when applied to a scheme demanding infinite precision.

Kieu actually brings in the idea of a small perturbation into his scheme, and it is instructive to see how it illustrates this general point. He introduces it because of a technical difficulty. This is that in general, the least value of the polynomial being considered will not occur for a unique argument. For instance, the example given above will yield a minimum value of 1 at (2,3), (5,7), (12, 17) and infinitely many more values. This degeneracy affects the adiabatic theory and means that the system will not evolve into a suitable state for the scheme to work. Kieu's answer (Kieu 2003d, p. 4) is to apply a small perturbation to the system which will separate the eigenstates. He cites this as a familar technique in quantum chemistry for resolving degenerate energy levels. But in quantum chemistry, energy levels with unboundedly large values do not play a crucial role, and the concept of 'small' perturbation makes sense. In Kieu's scheme the 'small' perturbation parameter will still give unboundedly large perturbations when applied to the unboundedly large energy states which are crucial to the method. It may be that reformulation and correction of Kieu's argument may eliminate the 'degeneracy' problem, and so it may not in itself be a fatal flaw. But Kieu's treatment of it illustrates how this author has not seriously considered how an infinitely precise model must be treated quite differently from one in normal applied mathematics.

7. **Contrast with Penrose's argument:** How does Kieu's theory compare with the speculative theory of Penrose (1989, 1994) that there are uncomputable effects in quantum mechanics? It is completely different. Again, it is important to distinguish the roles of the unitary evolution and the reduction/measurement/collapse parts of quantum mechanics. Penrose suggests that there is a hidden structure to the latter: that reduction is not actually random but rests on some quite unknown dynamical principle which involves uncomputability. Penrose is not *applying* the rules of quantum mechanics as presently understood, but proposing that they must be *superseded* by new ones with greater content. These unknown laws might be quite different from differential equations, possibly involving boundary conditions of some quite new type. They might involve the non-locality of wave-functions in an essential way, and might not even be expressible in terms of space and time, requiring a deeper geometrical setting such as twistor space. Since we have little idea of what these laws might be, questions of precision and approximation cannot even be framed.

Another difference is this: if Kieu were right, then one could (in principle) build a quantum processor that would act as an oracle for all polynomials: one could type in a polynomial equivalent to asking for the truth of the Fermat-Wiles theorem, or the Riemann hypothesis, and in a finite time obtain the (probably) correct answer. Penrose does not propose any such thing, and indeed emphasises that the uncomputability involved in his theory must be something 'completely different' from such an oracle.

**8. Connection with the Church-Turing thesis:** The interest of this topic arises because of its theoretical relationship with the bounds on classical Turing computability. The significance of these bounds is often referred to in terms of the 'Church-Turing thesis', and so there is much argument about what Kieu's proposal has to say about the truth or otherwise of this thesis. Unfortunately, there is no unanimity regarding what the Church-Turing thesis actually asserts. The philosophers Copeland and Proudfoot (1999) insist that Church and Turing only discussed the calculations undertaken by a human being applying some rule, and said nothing at all about physical machines. In contrast, the computer scientist Andrew Chi-Chih Yao (2003) takes it to refer to 'any conceivable hardware system' effecting a calculation, commenting that this 'may not have been what Church and Turing believed,' but that it represents a commonly accepted view. Because of this uncertainty in definition, some writers distinguish a 'physical' Church-Turing thesis as opposed to a Church-Turing thesis confined to mathematical logic.

In fact, Copeland and Proudfoot are mistaken, for Church (1937) explicitly characterised computable numbers as those which could be calculated by 'a computing machine, occupying a finite space and with working parts if finite size', referring to 'a human calculator' only as a particular example of such a machine. Turing likewise wrote freely of 'machines' without making the distinction claimed as all-important by Copeland and Proudfoot. Thus Yao better represents Church and Turing's general thrust, but his definition is rather too loose, as it lacks the finiteness condition carefully imposed by Church. Such a condition is necessary: one could 'conceive of' an infinite register, embodied in an infinite universe, set up as an infinite crib-sheet which could be trivially read off to 'solve' Hilbert's Tenth Problem.

Naturally, these remarks on history and conventional definitions make no difference to the actual science and mathematics of a scheme such as Kieu's. But we should certainly take due note of the original expression 'effective calculability' used for the concept that Turing elucidated with his Turing machine definition. 'Effective' means something that can actually be done. If a 'broad' concept of calculation is adopted which allows anything that can be merely postulated, then anything is calculable. (To compute X, postulate a machine with the property of printing X). The role of infinite quantities is crucial here. Waiting for an infinite time is not counted as effective. An infinite register of data must be ruled out likewise. An 'accelerated Turing machine' eliminates infinite time, but at the cost of demanding unboundedly short time-steps. Given the breakdown of space-time structure below the level of the Planck time, requiring such unboundedly short times would appear to be as ineffective as an infinite waiting time. Likewise an infinite data store can be compressed into a finite space at the cost of requiring an unboundedly precise measurement, something equally ineffective. A particularly trivial example of the latter trade-off, relying on measuring 'an amount of electricity' to infinite precision, is advanced by Copeland and Proudfoot (1999) as a model of how the hypercomputer revolution may take off. Another important finiteness property of Turing machines is that only a finite number of squares is used in the course of a calculation. But Kieu's procedure requires an infinite number of quantum states to be used, moreover with unbounded speeds.

Rather than take a dogmatic view of the Church-Turing thesis, it is probably better to expect it to evolve in conjunction with our better understanding of physical law, with Church's finiteness conditions refined so as to apply in contexts where the classical concepts of 'size' and 'parts' are inadequate. Meanwhile, if proponents of 'hypercomputing' schemes wish to argue that unbounded speeds of operation, unboundedly small components, infinitely many working states or other such properties should be counted as 'effective' aspects of procedures, then their argument should be made explicit, rather than buried in the formalism, and some indication should be given of how these infinitudes could be physically realized.

**9. Conclusion:** Can quantum computing solve classically unsolvable problems? The general considerations above indicate that nothing like Kieu's scheme will constitute an effective calculation. But Kieu's proposal does have mathematical interest and may well help to stimulate new work. It draws attention to quantum mechanical procedures, neglected by logicians such as Gandy and Sieg. (Indeed the logicians' restricted framework does not even allow for procedures using quantum entanglement, such as are already being used in quantum cryptography.) In particular, Kieu draws attention to the 'adiabatic cooling' method in which there are no discrete 'steps' of calculation. This again illustrates how existing analyses of 'machines' are over-restrictive. (It is worth noting that Church (1937) did not actually mention discrete steps, although logicians generally assume that 'effective' necessitates a finite number of operations). Thus Kieu's flawed work may be helpful in stimulating further analysis and a better understanding of the relationship between computability and physics.

**Acknowledgements:** This analysis was stimulated first by correspondence with Martin Davis, and then by an invitation to speak at the conference of the European Society for Analytical Philosophy in August 2005. Warren D. Smith's analysis has also been of great value.